\documentclass[twocolumn,showpacs,preprintnumbers, reprint, amsmath,
amssymb,prl]{revtex4-1}

\usepackage{graphicx}% Include figure files
\usepackage{dcolumn}% Align table columns on decimal point
\usepackage{bm}% bold math

\graphicspath{{./figs/}}

\begin{document}
\title{Lift-off dynamics in a simple jumping robot}
\author{Jeffrey Aguilar}
\affiliation{School of Mechanical Engineering, Georgia Institute of Technology,
Atlanta, Georgia 30332, USA}
\author{Alex Lesov}
\affiliation{School of Physics, Georgia Institute of Technology,
Atlanta, Georgia 30332, USA}
\author{Kurt Wiesenfeld}
\affiliation{School of Physics, Georgia Institute of Technology,
Atlanta, Georgia 30332, USA}
\author{Daniel I.\ Goldman*}
\affiliation{School of Physics, Georgia Institute of Technology,
Atlanta, Georgia 30332, USA}
\date{\today}

\begin{abstract}

We study vertical jumping in a simple robot comprising an actuated mass-spring arrangement.  The actuator frequency and phase are systematically varied to find optimal performance.  Optimal jumps occur above and below (but not at) the robot's resonant frequency $f_0$.  Two distinct jumping modes emerge: a simple jump which is optimal above $f_0$ is achievable with a squat maneuver, and a peculiar stutter jump which is optimal below $f_0$ is generated with a counter-movement. A simple dynamical model reveals how optimal lift-off results from non-resonant transient dynamics.

\end{abstract}

\maketitle
{\em Introduction} -- Organisms~\cite{alexanderbook,Dickinson2000a} and robots~\cite{cheAumb09,plaAbue06,Pfeifer2007} that inhabit the terrestrial world must run, crawl, and jump over a diversity of substrates and do so by effective deformations of appendages and bodies. While both simple ~\cite{LABid2600119, kuoAdon} and complex~\cite{zajac93} models have been created to study optimal movement patterns, simple models have the ability to be fully analyzed and can thus provide guidance for simplifying control of more complex devices, and even reveal principles of biological locomotion~\cite{holAful06}.

Jumping is an important behavior for animals and robots and is interesting, since it involves a transient burst of activity. Biological studies have revealed mechanisms of jumping in a diversity of organisms \cite{Harris2002,Aerts1998,Lutz1994,Kubo1999,zajac93}. In robotics, biologically inspired legged jumping robots have been constructed as an alternative to wheeled robots to better traverse rough terrain \cite{Armour2007,Kovac2008,Babic2009,Hyon2002,Niiyama2007}. The initial movement strategies for optimal jumping are typically chosen by empirical tuning for steady state hopping \cite{Raibert1985,Hyon2002,Ohashi2006} or squat jumps \cite{Niiyama2007,Niiyama2008,Babic2009}. Systematic studies of the dynamics of transient behaviors, critical to issues of lift-off, are relatively scarce.

In this paper we perform a detailed study of a simple jumping robot, a 1D mass-spring system with an actuated mass; this model was originally developed as a template for hopping in steady-state\cite{Full1999} and encapsulates the leg compliance and an organism's ability of self-deformation via leg actuation. Systematic variation of forcing parameters reveals complex dynamics which are sensitive to amplitude, phase and frequency.  Contrary to our initial expectation, optimal jumping does not occur at resonance. We introduce a reduced ``piecewise linear" equation of motion for the robot to analyze the transient dynamics; another non-linearity appears as a result of ground collisions. The model reveals a richness in behavior governed by interplay of forced and free motion.

\begin{figure}[h]
\begin{centering}
\includegraphics[width={1\hsize}]{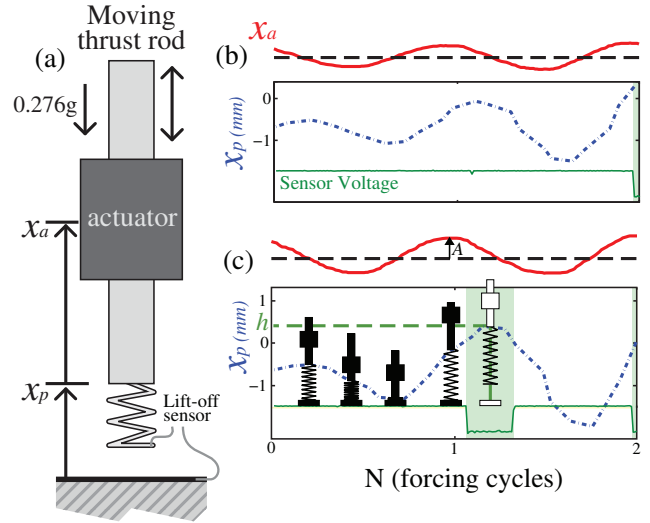}
\caption{Jumping robot (a) schematic diagram. (b,c) Actuator position $x_a$ and video tracked position of the thrust rod $x_p$, and the sensor voltage for $A = 0.19$ mm in which robot does not lift-off and $A=0.30$ mm in which the robot lifts off.}
\end{centering}
\end{figure}

{\em Experiment and model} -- The robot (total mass $m = 1.18$ kg) consisted of a linear motor actuator (Dunkermotoren ServoTube STA11) with a series spring ($k=5.8$ kN/m) rigidly attached to the bottom end of the actuator's lightweight thrust rod. The actuator was mounted to an air bearing which allowed for 1D, and nearly frictionless, motion. Due to power limitations in the actuator, the bearing was inclined at  $15\,^{\circ}$ relative to the horizontal, reducing gravitational acceleration to $0.276g$. The position of the actuator relative to the bottom of the thrust rod, $x_{a}$, was controlled such that $x_{a}(t)=A\,\sin(2 \pi f t + \phi)$, where the amplitude, $A$, frequency, $f$, and initial phase offset, $\phi$ are constant during a jump. The natural frequency of the robot, $f_{0}=\frac{1}{2 \pi} \sqrt{k/m}= 11.13$ Hz. Video tracking of the decay of the free oscillations of the robot (with fixed actuator position and spring on the ground) indicated that damping was small (damping ratio, $\zeta \approx 0.01$), and thus the resonant frequency was virtually equal to the natural frequency.

The jumping platform consisted of a rigid aluminium plate; the coefficient of restitution of the robot with the plate was $0.8 \pm 0.06$ over the range of relevant collision speeds. To detect lift-off, a continuity sensor attached to the bottom of the metal spring measured an open circuit when the spring left the ground. Time to lift-off from the onset of actuator activation and time of flight were determined to 1 msec and jump height was calculated from time of flight.

In concert with the experiments, we studied a simple dynamical model of the robot, with equation of motion
\begin{equation}
\ddot{x}_{p}=-\ddot{x}_{a}\frac{{m_{a}}}{m}-\alpha\left(\dot{x}_{p}\frac{c}{m} + x_{p}\frac{k}{m}\right) -g
\end{equation}
where $k$, $c$, and $g$ are stiffness, damping and gravity, respectively. The total mass $m$ is the sum of the actuator/air-bearing mass, $m_{a}=1.003$ kg, and the mass of the thrust rod, $m_p=0.175$ kg. The piece-wise constant, $\alpha = 1, \mbox{ if } x_p<0$ and $0, \mbox{ if } x_p \geq{0} $. A constant coefficient of restitution of $0.8$ (measured from experiment) modeled the collision of the spring with the ground.

\begin{figure}[h]
\begin{centering}
\includegraphics[width={1\hsize}]{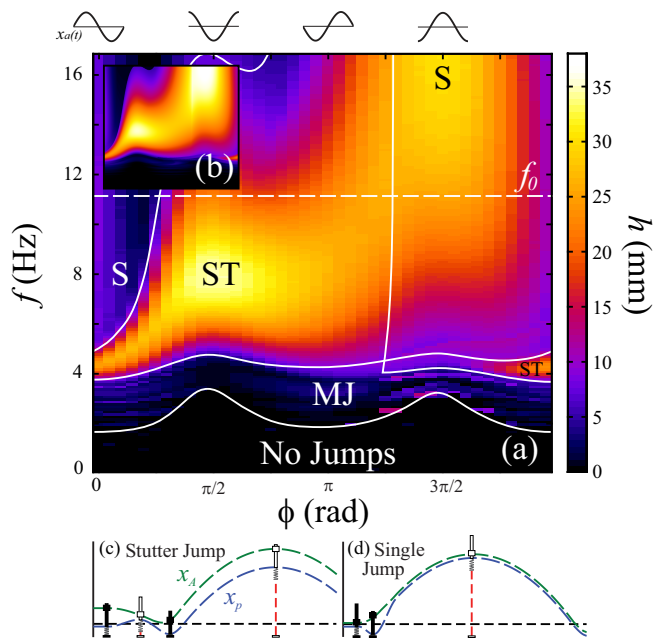}
\caption{Jump height and jumping modes. (a) Experimental jump height, $h$, as a function of $f$ and $\phi$ with illustrated actuator trajectories (top) at different $\phi$ with $A=4$ mm. White lines (derived from model) separate different jumping modes, with ST indicating stutter jump, S for single jump, and MJ for multi-jump. (b) inset shows model (Eq. 1) with same parameters. (c,d) illustrate trajectories for the stutter and single jumps. Robot is airborne (white robots) when the rod position, $x_{p}>0$. Global actuator position, $x_{A}$, not to scale with rod length.}
\end{centering}
\end{figure}

{\em Lift off and jump height} -- In all experimental runs, at $t=0$ the actuator was commanded to move from rest (Fig. 1b,c). For fixed frequency, below a minimum amplitude $A_{min}(f)$, no lift-off was detected and above $A_{min}(f)$ the robot was able to jump; $A_{min}(f)$ was determined by an iterative procedure in which a binary search was implemented until $A_{min}(f)$ was determined to within $0.00625$ mm, the resolution of the actuator encoders. As expected, when the actuator was continuously activated, the absolute minimum of $A_{min}$ over all frequencies occurred at the resonant frequency $f_0$, and was independent of $\phi$. However, since we were interested in rapid jumps from rest, actuator forcing was then restricted to only one cycle ($N=1$). To our surprise, things were qualitatively different: the smallest $A_{min}(f)$ did not occur at $f_0$, and varied with phase offset $\phi$.

We next systematically examined jumping height for $N=1$. We fixed $A=4$ mm, which was above $A_{min}(f)$ for $f>3.5$ Hz, and studied how jump height $h$ depended on $f$ and $\phi$. The mean $h$ from three trials was recorded. Variation in $h$ from jump to jump was small; the standard deviation of $h$ was less than 0.5 mm, or approximately 1\% of mean $h$, past approximately 4 Hz. Certain fractions of $f_0$ below 4 Hz exhibited significant variance due to small multi-jumps that occurred as a result of sub-resonant harmonics (max heights seen in MJ section of Fig. 2a). These frequencies were perfectly timed to allow multiple complete oscillations during motor actuation to precede a larger final jump.

Fig 2 shows the results of $6720 \times 3$ experiments. For fixed $\phi$, above a critical $f$ the robot was able to lift-off. Two broad maxima in $h$ were observed, neither occurring at $f_0$. Integration of Eq. 1 quantitatively reproduced the experiments, see Fig. 2b inset.

The two local maxima correspond to two distinct modes of jumping: a ``single jump" and a ``stutter jump". In the single jump mode, the robot compressed the spring and was propelled into the air. In the stutter jump mode, the robot performed a small initial jump followed by a larger second jump, see Fig. 2c. We used the model to determine the boundaries of the regions of the $\phi-f$ plane of the different modes. For large $\phi$ single jumps predominate while stutter jumps occurred at lower $f$ and $\phi$.

The emergence of the stutter jump was unexpected.  To understand its presence, consider (for example) the case $\phi=\pi/2$, so that the initial actuator acceleration is negative.  This causes the less massive thrust rod to be accelerated upward before moving down to compress the spring and then lift off again. Interestingly, the stutter jump was observed even for phases somewhat larger than $\pi$ (Fig 2), for which the initial actuator acceleration is expected to progress positively from 0. The reason lies in the physical constraint that the actuator must start from rest, regardless of phase offset. Thus, any phase offset corresponding to a non-zero initial actuator velocity causes an initial impulse acceleration ({\em i.e.} the initial actuator trajectory is not an ideal sine wave). For a phase such as $\pi$, the initial, brief actuator acceleration is large and negative, causing an intermediate hop.

We next examined various characteristics associated with optimal jump height, see Fig. 3. At the optimal $f$ of each phase $\phi$, maximum $h$ was determined and displayed two broad maxima (Fig 3a); the maximum $h$ values were insensitive to $f$ and were nearly $10 \times$ larger than $A$. The $f$ which gave maximum $h$ was less than $f_0$ for stutter jumps and greater than $f_0$ for single jumps (Fig 3b). The time to lift-off (Fig. 3c) was smaller for single jumps than stutter jumps. As we will show below, peak power expended in deforming the system $P_{def}$ scaled like $f^3$ and thus increased dramatically for single jumps.

\begin{figure}[h]
\begin{centering}
\includegraphics[width={.95\hsize}]{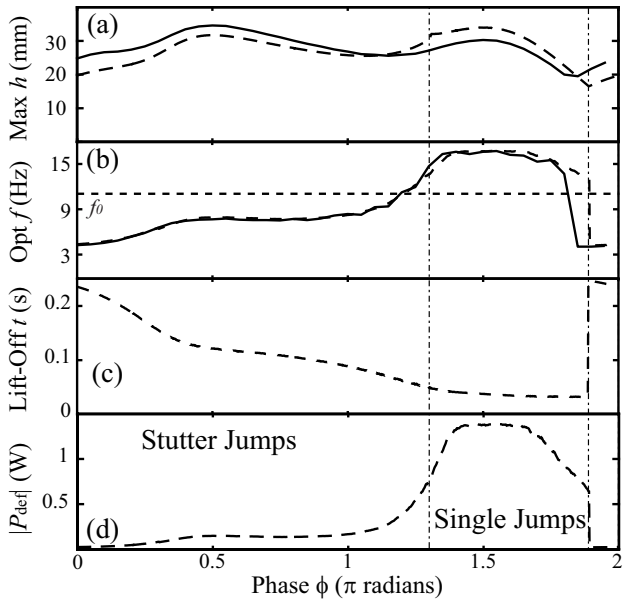}
\caption{Experimental (solid) and simulation (dashed) results of (a) maximum jump height at each phase, (b) the corresponding optimal frequency for each phase, (c) time to lift-off, and (d) deformation power at optimal frequency; initial transients are omitted in this calculation.}
\end{centering}
\end{figure}

{\em Theory of transient mixing} -- At first glance, Eq.(1) looks completely tractable.  Unfortunately, the discontinuity associated with the factor $\alpha$ renders the equation ``piecewise linear", which is to say nonlinear.  Indeed, simulations of Eq. 1 show a wide variety of behaviors (including bifurcations, hysteresis, chaos). The situation is reminiscent of other piecewise linear dynamical systems which display complex dynamics, including the tent map \cite{gucAhol} and the bouncing ball \cite{tufAmel}.  Nevertheless, using analysis and numerics, Eq. 1 allows us to gain insight into the experimental observations.  We are particularly interested in why optimal jumps occur only off resonance.

Consider first the peak labeled S in Fig. 2a, representing the highest single jumps.  This peak occurs at actuator phases near $\phi=3\pi/2$.  For a relatively low thrust rod mass, jump height is proportional to the square of the absolute actuator velocity, $\dot{x}_{A}(t)=\dot{x}_{p}+\dot{x}_{a}$, at take-off. Neglecting damping and collisional loss, at $\phi=3\pi/2$, this velocity is (solving Eq. 1 with $\alpha = 1, c=0, m_a=m$)
\begin{align}
\dot{x}_{A}(t)=\frac{2 \pi A f^2}{f_{0}^2-f^2}f_{0}\left(-\sin{2 \pi f_{0}t}+\left(f_{0}/f\right)\sin{2 \pi f t}\right)
\end{align}

The take-off velocity is thus a prefactor times the sum of two sinusoids (the one at frequency $f_0$ represents the transient response, which mixes with the steady state contribution).  The prefactor generally favors $f$ near $f_0$, but destructive interference suppresses $\dot{x}_A$ too close to resonance.  Moving off resonance, the prefactor favors higher $f$ over lower, so the optimum $f$ lies somewhat above $f_0$. This argument holds regardless of $A$.

Understanding the optimality of the stutter jump is more complicated.  The key is to consider the system energetics, and in particular the conditions that maximize the total work done during the drive cycle.  The instantaneous power input is $P = F_{ext} v_{m}$, where $F_{ext}$ is the total external force (including gravity and spring forces) and $v_m$ is the center-of-mass robot velocity, which is to good approximation the absolute actuator velocity, $\dot{x}_{A}$. The total work done by external forces is maximized when $\dot{x}_{A}$ both (1) is large in magnitude, and (2) has the same sign as the $F_{ext}$.

\begin{figure}[h]
\begin{centering}
\includegraphics[width={.9\hsize}]{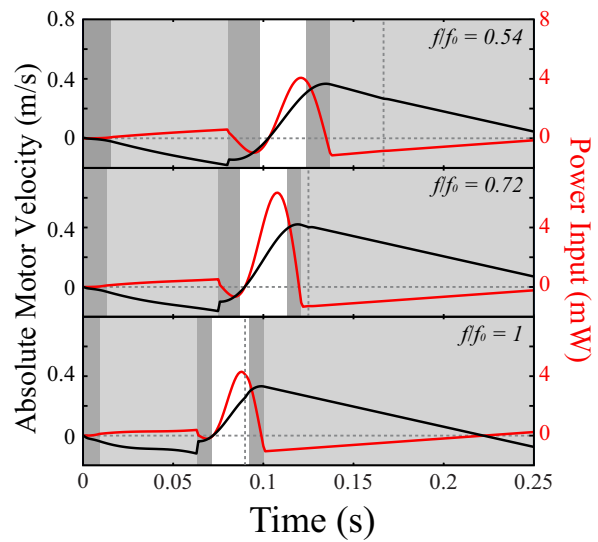}
\caption{Simulated time trajectories of absolute actuator velocity (black) and power input by external forces (red) for $f/f_{0} = 0.54, 0.72$ (optimal), and 1, at $\phi = \pi/2$. The vertical dotted line indicates when the actuator stops. Light gray areas: aerial state ($x_{p}>0$); dark gray: negative force ground state ($mg/k<x_{p}\leq0$); white: positive force ground state ($x_{p}\leq(mg/k)$).}
\end{centering}
\end{figure}

The situation is illustrated in Fig. 4, for $\phi=\pi/2$, {\it i.e.} when the stutter jump is most effective. At $f=f_{0}$ (lower panel), the actuation is too fast and the actuator turns off well before lift-off.  At a low $f$ (top panel), the actuation is too slow:  the actuator turns off well after lift-off, so much of the power stroke is wasted.  The optimal drive (middle panel) lies somewhere in between.  In addition, an optimal stutter jump depends not only on the phasing of competing sinusoids while on the ground (just as for single jumps), but also on the proper timing of ground and aerial states.  The latter varies with $\phi$ and does not generally occur at $f_{0}$. This sensitivity to proper timing explains the narrow frequency bandwidth required to achieve optimal jump heights using the stutter jump mode.  A further consequence is a strong dependence of optimal $f$ with respect to $A$: larger $A$ produce lower optimal $f$, and smaller $A$ produce higher optimal $f$.  In contrast, the optimal $f$ for the single jump mode does not show a strong dependence on $A$.

Assuming small damping and no collisional losses, deformation power (defined as $P_{def} = m_{p}\ddot{x}_{p}\dot{x}_{p} + m_{a}\ddot{x}_{A}\dot{x}_{A}$) was calculated as $P_{def}=\frac{4 \pi^3 m_a m_p}{m} A^2 f^3$. Thus the stutter jump is energetically advantageous since it has a lower optimal $f$ than the single jump. In fact, the stutter jump uses nearly an order of magnitude less power to achieve comparable jump height to the single jump.

{\em Conclusion}-- We have analyzed the dynamics of lift-off in the simplest hopping robot and found that the performance is quite rich and remarkably sensitive to starting phase trajectory, largely a result of the transient dynamics in a linear-mass spring system. Unlike in steady state, in the transient regime, optimization occurs at non-resonant frequencies. The system becomes hybrid for certain parameters as a stutter jump emerges. This mode achieves comparable jump height but uses less power. Analysis of a simplified model reveals that impulse accelerations and discontinuous transitions to the aerial state are essential ingredients in understanding the dynamics.

Our model provides insight which more complex and multi-functional robots can use to execute rapid jumps and starts. Biologically, our model is in accord with a previous model of bipedal jumping which predicted that counter-movement achieves greater jump height than the squat jump~\cite{Alexander1995}. A quick single jump that resembles a squat jump is beneficial when a fast escape is essential, while a slower stutter jump similar to a counter-movement can achieve comparable jump height. Primates like Galagos (bushbabies) have been documented to perform this double jump behavior to reach a higher platform \cite{Gunther1991}. Based on the power arguments above, we hypothesize that this mode is advantageous.

It would be interesting to investigate how other factors, intrinsic and environmental, affect optimal performance.  A non-sinusoidal actuation could improve jump height, take off time, or efficiency.  Animals jump off compliant surfaces (like tree branches) and from deformable substrates (like sand). Systematic studies of the jumping robot under analogous conditions could yield insight into optimal strategies in these and other (man-made) actors.

{\em Acknowledgements--} We thank Harvey Lipkin, Paul Umbanhowar, Nick Gravish, and Yang Ding for discussion, as well as Andrei Savu for apparatus construction. This work was supported by the GEM Fellowship, the Burroughs Wellcome Fund and the ARL MAST CTA.
\bibliographystyle{apsrev4-1}

\end{document}